\documentclass[12pt,a4paper]{article}
\usepackage{amsmath,amssymb,bm,ascmac}
\usepackage[dvipdfmx]{graphicx}
\usepackage{color}
\usepackage{authblk}
\begin{document}
\title{AdS/CFT and local renormalization group with gauge fields
}
\author[1]{Ken Kikuchi}
\author[1,2]{Tadakatsu Sakai}
\affil[1]{Department of Physics, Nagoya University}
\affil[2]{KMI, Nagoya University}
\maketitle

\makeatletter
\renewcommand{\theequation}
{\arabic{section}.\arabic{equation}}
\@addtoreset{equation}{section}
\makeatother

\newcommand{\defi}{\stackrel{\mathrm{def}}{\iff}}
\newcommand{\tdif}[2]{\frac{d#1}{d#2}}
\newcommand{\pdif}[2]{\frac{\partial #1}{\partial #2}}
\newcommand{\ddif}[2]{\frac{\delta #1}{\delta #2}}
\newcommand{\map}[1]{\stackrel{#1}{\rightarrow}}
\newcommand{\longmap}[1]{\stackrel{#1}{\longrightarrow}}
\newcommand{\fsl}[1]{\not\!#1}
\newcommand{\ld}[1]{\mathfrak L_{#1}}
\newcommand{\dk}[2]{\frac{d^{#1}#2}{i(2\pi)^{#1}}}
\newcommand{\dz}[1]{\frac{d#1}{2\pi i}}
\newcommand{\dv}[2]{d^D#1\sqrt{-#2}}
\newcommand{\delB}{\bm\delta_\mathrm{B}}

\newcommand{\nn}{\nonumber}
\newcommand{\cD}{\mathcal D}
\newcommand{\fD}{\mathfrak D}
\newcommand{\cJ}{\mathcal J}
\newcommand{\cL}{\mathcal L}
\newcommand{\cW}{\mathcal W}

\newcommand{\nabb}{\bm\nabla}

\newcommand{\mh}{\hat{\mu}}
\newcommand{\nh}{\hat{\nu}}
\newcommand{\rhoh}{\hat{\rho}}
\newcommand{\Ah}{\hat{A}}

\thispagestyle{empty}
\setcounter{page}{0}

\begin{abstract}
We revisit a study of local renormalization group (RG) 
with background gauge fields incorporated using
the AdS/CFT correspondence.
Starting with a $(d+1)$-dimensional bulk gravity coupled to
scalars and gauge fields, we derive a local RG equation
from a flow equation by working in the Hamilton-Jacobi
formulation of the bulk theory.
The Gauss's law constraint associated with gauge symmetry plays an
important role.
RG flows of the background gauge fields are governed by
vector $\beta$-functions, and some interesting properties
of them are known to follow.
We give a systematic rederivation of them on the basis of
the flow equation.
{}Fixing an ambiguity of local counterterms in such a manner 
that is natural from the viewpoint of the flow equation,
we determine all the coefficients uniquely appearing
in the trace of the stress tensor for $d=4$.
A relation between a choice of schemes and a Virial current
is discussed.
As a consistency check, these are found to satisfy 
the integrability conditions of local RG transformations.
{}From these results, we are led to a proof of a
holographic $c$-theorem by finding out a full family
of schemes where a trace anomaly coefficient is related
with a holographic $c$-function.
\end{abstract}

\setcounter{section}{+0}
\setcounter{subsection}{+0}

\newpage

\section{Introduction}
Coupling `constants' are literally regarded as constants 
in ordinary quantum field
theories (QFTs). However, it is an interesting question 
to ask what happens when
they have spacetime dependence. 
A method called local renormalization group
(RG) puts this idea in practice. 
That is, we lift spacetime independent coupling
constants $g$ to spacetime dependent coupling functions 
$g(x)$ \cite{O1991,JO}.
The coupling functions can be regarded as external fields. 
Correlation functions are thus obtained by functional derivatives 
of the generating functional $\Gamma[g(x)]$ of
connected graphs (a.k.a. the Schwinger functional) with respect to $g(x)$. 
This rule is called the Schwinger's action principle \cite{S51}. 
In the AdS/CFT correspondence \cite{Maldacena}, 
$\Gamma[g(x)]$ is identified with an on-shell action of a bulk
gravity dual with the external fields corresponding to 
boundary values of bulk fields \cite{GKPW}.
Studies of RG flows using the AdS/CFT correspondence have been
done extensively so far. In particular, an analysis in
this line using a flow equation was
first made by de Boer, Verlinde and Verlinde \cite{dVV}, 
and then generalized in \cite{FMS1}. For a review, see
\cite{FMS;review}.
In these papers, 
bulk gravity theories coupled to scalar fields with
a generic metric were investigated.
It was revealed that the flow equation of a bulk gravity
yields a local RG equation of $\Gamma[g(x)]$ with
the Weyl anomalies of the boundary QFT reproduced 
correctly in this framework.
{}For recent developments in local RG, see also
\cite{OS, BKRV,Auzzi, Raj}.
One of the purposes of this paper is to generalize these results 
by introducing gauge fields in the bulk side. 
This is partially motivated by a desire to  bring a somewhat 
mysterious quantity called a vector  $\beta$-function 
\cite{O1991, JO} to light. 
This characterizes how background gauge fields coupled to
currents flow under RG transformations.
Some of the properties mentioned above were derived from
the Wess-Zumino consistency conditions concerning
local RG transformations. A paper \cite{N13} found that
the AdS/CFT correspondence explains these properties nicely.
In this paper, we make a systematic rederivation of these 
results on the basis of the flow equation.
One of the advantages of our analysis throughout this paper 
is to clarify scheme dependence of the results, that is,
how they are affected by a choice of local counterterms.
In particular, we point out a close relationship
between a choice of schemes and a Virial current.

The organization of the paper is as follows. 
In section 2, we formulate a $(d+1)$-dimensional gravity
dual model in the Hamilton-Jacobi formalism. 
We derive the flow equation and give some
comments on its general aspects. Some suggested properties of 
the vector $\beta$-functions such as (i) gradient property, 
(ii) orthogonality, (iii) Higgs-like relation between anomalous 
dimensions and (iv) the relation between a vanishing vector
$\beta$-function and non-renormalization of current operators can be readily
confirmed. In section 3, we perform explicit calculations in $d=4$. 
We obtain closed expressions of anomaly coefficients including 
central charges
\cite{c-thm,Cardy,a-thm}
with an emphasis on a scheme choice we made.
We can see the coefficient functions satisfy integrability conditions 
that come from the Wess-Zumino consistency condition associated
with the local RG transformation \cite{O1991,JO}. 
{}For an earlier work in this line, a paper \cite{Raj}
discusses five-dimensional bulk gravity coupled to scalar fields 
without gauge fields.
It is shown there that 
the anomaly coefficients computed in that model 
satisfy the Wess-Zumino consistency conditions.
As an application of our results, a holographic $c$-theorem in 
$d=4$ dimensions is proven 
by finding out a full family of schemes 
where a monotonically decreasing function
under RG flows can be constructed from an anomaly coefficient.
This is an extension of a paper \cite{Erd01}.
We collect our notations, some useful formulae and lengthy equations 
in appendices.

\section{Formalism}
We start with a bulk action in $(d+1)$-dimensions:
\begin{align}
\mathbf S&\left[\hat\gamma_{\mh\nh}(x,\tau),\hat\phi^I(x,\tau),\Ah_{\mh}(x,\tau)
		\right]\nonumber\\
=&\int_{M_{d+1}}d^{d+1}X\sqrt{\hat\gamma}\,\Big\{V(\hat\phi)-\hat R_{(d+1)}
+\frac12L_{IJ}(\hat\phi)\hat\gamma^{\mh\nh}\hat\nabb_{\mh}\,\hat\phi^I
\hat\nabb_{\nh}\,\hat\phi^J
+\frac14J(\hat\phi)\hat F^a_{\mh\nh}\hat F^{a\mh\nh}\Big\}\nonumber\\
&-2\int_{\Sigma_d}d^dx\sqrt{\hat h}\,\hat K \ .
\label{bulkS}
\end{align}
Here $\hat\gamma_{\mh\nh}$ denotes a bulk metric.
Using ADM decomposition, it becomes
\begin{align}
  ds^2=\hat\gamma_{\mh\nh}\,dX^{\mh}dX^{\nh}
=\hat N^2(x,\tau)d\tau^2+\hat h_{\mu\nu}
(x,\tau)(dx^\mu+\hat\lambda^\mu(x,\tau)d\tau)(dx^\nu+\hat\lambda^\nu(x,\tau)
d\tau) \ ,
\end{align}
where $\hat h_{\mu\nu}$ is an induced metric on a $d$-dimensional hypersurface
$\displaystyle{\Sigma_d:=\{X\in M_{d+1}|\tau=\text{const.}\}}$.
We have 
$\hat\gamma=-\det(\hat\gamma_{\mh\nh}),\hat h=-\det(\hat h_{\mu\nu})$.
On each slice,
an extrinsic curvature is defined as
\begin{align}
\hat K_{\mu\nu}:=\frac1{2\hat N}(\partial_\tau\hat h_{\mu\nu}
-\hat\nabla_\mu\hat \lambda_\nu-\hat\nabla_\nu\hat \lambda_\mu) 
\ ,
\end{align}
and
$\hat K:=\hat h^{\mu\nu}\hat K_{\mu\nu}$.
The hatted quantities mean off-shell without the equations
of motion imposed.
The covariant derivatives are defined as
\begin{align}
\hat \nabb_{\mh}\,\hat\phi^I:=
\hat\nabla_{\mh}\,\hat\phi^I-i\hat A^a_{\mh}\,(T^a\hat\phi)^I
\ . 
\end{align}
Here $\hat\nabla_{\mh}$ denotes a covariant derivative
associated with the Levi-Civita connection, $\hat\Gamma^{\mh}_{\nh\rhoh}$\,,
constructed from
$\hat\gamma_{\mh\nh}$.
$T^a$ is a generator of the gauge group $G$.
Since we want to recognize
$\phi$ as real coupling functions, we restrict the symmetry $G$ to a group which
has real representations such as $SO(N)$.
{}For details, see Appendix \ref{notation}.

Before proceeding, some comments on earlier works on the 
AdS/CFT correspondence with bulk gauge fields are in order.
This was first investigated by a paper \cite{Bia} 
on the basis of holographic renormalization, and since then
has been discussed extensively from many perspectives.
A systematic algorithm for solving the HJ equations in the
cases with Abelian gauge fields and neutral scalar fields coupled 
is obtained in \cite{Papa05}.
Using this algorithm, 
a paper \cite{Lind} studies a class of AdS/CMP models that are
parametrized with a Lifshitz exponent $z$.
In this section, we aim to give a formalism
for analyzing more general systems of bulk gravity coupled with
nonAbelian gauge fields and charged scalar fields
by extending a flow equation that was first developed in
\cite{dVV}.
Many of the results we give below coincide with those
found already in the papers quoted above.

Working in a Hamilton formalism with $\tau$ regarded as
a time direction rewrites the action (\ref{bulkS}) 
in a first-order form:
\begin{align}
\mathbf S&
=\int d^dxd\tau
		\sqrt{\hat h}\Bigg\{\hat\pi^{\mu\nu}\partial_\tau\hat h_{\mu\nu}
			+\hat\pi_I\partial_\tau\hat\phi^I+\hat\pi^{a\mu}\partial_\tau\hat A^a_\mu
				\nonumber\\
	&\hspace{20pt}+\hat N\Big[\frac1{d-1}({\hat\pi^\mu}_\mu)^2-\hat\pi_{\mu\nu}^2
		-\frac12L^{IJ}(\hat\phi)\hat\pi_I\hat\pi_J-\frac1{2J(\hat\phi)}\hat h^{\mu\nu}
			\hat\pi^a_\mu\hat\pi^a_\nu\nonumber\\
	&\hspace{60pt}+V(\hat\phi)-\hat R_{(d)}+\frac12L_{IJ}(\hat\phi)\hat h^{\mu\nu}
		\hat \nabb_\mu\hat\phi^I\hat \nabb_\nu\hat\phi^J+\frac14J(\hat\phi)
			\hat F^a_{\mu\nu}\hat F^{a\mu\nu}\Big]\nonumber\\
	&\hspace{20pt}+\hat\lambda^\mu\left[2\hat\nabla^\nu\hat\pi_{\mu\nu}
		-\hat\pi_I\hat \nabb_\mu\hat\phi^I-\hat F^a_{\mu\nu}\hat\pi^{a\nu}\right]
			\nonumber\\
	&\hspace{20pt}+\hat A^a_\tau\Big[{\hat \nabb^a}_{b\nu}\hat\pi^{b\nu}-i(T^a\hat\phi)^I
\hat\pi_I\Big]\Bigg\} \ .
\label{bulkS:1st}
\end{align}
Here the canonical momenta conjugate to 
$\hat h_{\mu\nu},\ \hat\phi^I$ and
$\hat A^a_\mu$ are respectively computed to be
\begin{align}
\hat\pi^{\mu\nu}:=&\pdif{\mathcal L_{d+1}}{(\partial_\tau\hat h_{\mu\nu})}
=\hat K^{\mu\nu}-\hat h^{\mu\nu}\hat K,
\ ,\\
\hat\pi_I:=&\pdif{\mathcal L_{d+1}}{(\partial_\tau\hat\phi^I)}=\frac1{\hat N}
L_{IJ}(\hat\phi)(\hat\nabb_\tau\hat\phi^J
-\hat\lambda^\mu\hat\nabb_\mu\hat\phi^J)
\ ,\\
\hat\pi^{a\mu}:=&\pdif{\mathcal L_{d+1}}{(\partial_\tau\hat A^a_\mu)}=\frac1
{\hat N^3}J(\hat\phi)\Big[\hat N^2\hat h^{\mu\nu}\hat F^a_{\tau\nu}
-\hat\lambda^\nu(\hat N^2\hat h^{\rho\mu}
+\hat\lambda^\rho\hat\lambda^\mu)\hat F^a_{\nu\rho}\Big]\ ,
\end{align}
with $L^{IJ}=L_{IJ}^{-1}$.
As evident, $\hat N$, $\hat\lambda^\mu$ and $\hat A_\tau$ are
the auxiliary fields, and their equations of motion 
yield the first-class constraints
\begin{align}
\hat H:=\frac1{\sqrt{\hat h}}\ddif{\mathbf S}{\hat N}
=&\frac1{d-1}({\hat\pi^\mu}_\mu)^2
-\hat\pi_{\mu\nu}^2-\frac12L^{IJ}(\hat\phi)\hat\pi_I\hat\pi_J
-\frac1{2J(\hat\phi)}\hat h^{\mu\nu}\hat\pi^a_\mu\hat\pi^a_\nu
\nonumber\\
&+V(\hat\phi)-\hat R_{(d)}+\frac12L_{IJ}(\hat\phi)\hat h^{\mu\nu}
\hat \nabb_\mu\,\hat\phi^I\hat \nabb_\nu\,\hat\phi^J+\frac14J(\hat\phi)
\hat F^a_{\mu\nu}\hat F^{a\mu\nu}\approx0 \ ,
\label{Hcon}
\\
\hat P_\mu:=\frac1{\sqrt{\hat h}}\ddif{\mathbf S}{\hat\lambda^\mu}
=&2\hat\nabla^\nu\hat\pi_{\mu\nu}-\hat\pi_I\hat \nabb_\mu\,\hat\phi^I
-\hat F^a_{\mu\nu}\hat\pi^{a\nu}\approx0 \ ,
\label{Momcon}
\\
\hat G^a:=\frac1{\sqrt{\hat h}}\ddif{\mathbf S}{\hat A^a_\tau}
=&\hat \nabb^a_{b\mu}
\hat\pi^{b\mu}-i(T^a\hat\phi)^I\hat\pi_I\approx0 \ .
\label{Gcon}
\end{align}
(\ref{Hcon}) and (\ref{Momcon}) are the Hamiltonian and momentum
constraints respectively that result from diffeomorphism in
the $(d+1)$-dimensional bulk spacetime. (\ref{Gcon}) is the Gauss's
law constraint due to the gauge symmetry $G$.

Suppose that we find a solution to the equations of motion of 
$\hat h_{\mu\nu}, \hat A_{\mu}$ and $\hat\phi^I$
with the constraints (\ref{Hcon}), 
(\ref{Momcon}) and (\ref{Gcon}) under a Dirichlet boundary
condition at $\tau=\tau_0$:
\begin{align}
\bar h_{\mu\nu}(x,\tau=\tau_0)=h_{\mu\nu}(x) \ ,~~
\bar A_{\mu}(x,\tau=\tau_0)=A_\mu(x) \ ,~~
\bar\phi^I(x,\tau=\tau_0)=\phi^I(x) \ .\nn
\end{align}
Here the bulk fields with a bar means on-shell.
Substituting the classical solutions into (\ref{bulkS}), 
we obtain the on-shell action as a functional of the
boundary values
\begin{align}
S[h_{\mu\nu}(x),\phi^I(x),A_\mu(x);\tau_0]
:=\int d^dx\int_{\tau_0}^\infty d\tau\sqrt{\bar h}\,\Big\{\bar\pi
		^{\mu\nu}\partial_\tau\bar h_{\mu\nu}+\bar\pi_I\partial_\tau\bar\phi^I
			+\bar\pi^{a\mu}\partial_\tau\bar A^a_\mu\Big\}
\ .
\label{onshellS}
\end{align}
{}Following the standard procedure in the Hamilton-Jacobi formalism,
it is verified that the variation of the on-shell action 
under the boundary values and the location of $\Sigma_d$
is given by
\begin{align}
\delta S[h(x),\phi(x),A(x);\tau_0]=-\int d^dx\sqrt{h}\,
\Big\{\bar\pi^{\mu\nu}
(x,\tau_0)\delta h_{\mu\nu}(x)+\bar\pi_I(x,\tau_0)\delta\phi^I(x)+\bar\pi^{a\mu}
(x,\tau_0)\delta A^a_\mu(x)\Big\} \ . 
\end{align}
We then obtain the Hamilton-Jacobi equations
\begin{align}
\bar\pi^{\mu\nu}(x,\tau_0)
=-\frac1{\sqrt{h}}\ddif S{h_{\mu\nu}(x)} \ ,~~
\bar\pi_I(x,\tau_0)=-\frac1{\sqrt{h}}\ddif S{\phi^I(x)}\ ,~~
\bar\pi^{a\mu}(x,\tau_0)&=-\frac1{\sqrt{h}}\ddif S{A^a_{\mu}(x)}\ ,~~
\frac{\partial S}{\partial\tau_0}=0 \ .
\end{align}
Inserting these into the Hamilton constraint (\ref{Hcon})
gives the flow equation
\begin{equation}
	\{S,S\}=\mathcal L_d \ ,
\label{floweq}
\end{equation}
where
\begin{equation}
\{S,S\}:=\left(\frac1{\sqrt{h}}\right)^2\left[-\frac1{d-1}\left(h_{\mu\nu}\ddif S
		{h_{\mu\nu}}\right)^2+\left(\ddif S{h_{\mu\nu}}\right)^2+\frac12L^{IJ}(\phi)\ddif
			S{\phi^I}\ddif S{\phi^J}+\frac1{2J(\phi)}h_{\mu\nu}\ddif S{A^a_\mu}\ddif S
				{A^a_\nu}\right]
\ ,
\label{SS}
\end{equation}
and
\begin{equation}
\mathcal L_d:=V(\phi)-R_{(d)}+\frac12L_{IJ}(\phi)\nabb^\mu\phi^I
\nabb_\mu\phi^J
		+\frac14J(\phi)F^a_{\mu\nu}F^{a\mu\nu}\ .
\label{Ld}
\end{equation}
Here
\begin{align}
\nabb_{\mu}\,\phi^I:=
\nabla_{\mu}\phi^I-i A^a_{\mu}\,(T^a\phi)^I
\ . 
\end{align}
$\nabla_{\mu}$ denotes a covariant derivative
associated with the Levi-Civita connection, $\Gamma^{\mu}_{\nu\rho}$\,,
constructed from the boundary metric
$h_{\mu\nu}$.

We show that the momentum constraint and the
Gauss's law constraint ensures $d$-dimensional diffeomorphism invariance 
and gauge invariance of the on-shell action, respectively.
{}First, we note that the Gauss's law constraint (\ref{Gcon}) and
the Hamilton-Jacobi equations give
\begin{align}
0&=\int d^dx\sqrt{h}\,\alpha^a\left({\nabb^a}_{b\mu}\pi^{b\mu}
-i(T^a\phi)^I\pi_I\right)
\nn\\
&=\int d^dx\left\{\nabb_\mu\alpha^a\ddif S{A^a_\mu}+i\alpha^a(T^a\phi)^I\ddif S{\phi^I}
		\right\}\label{Gaussopid}\\
&=\int d^dx\left(\delta_\alpha^{\rm gauge} A^a_\mu\ddif S{A^a_\mu}
+\delta_\alpha^{\rm gauge}\phi^I\ddif S{\phi^I}\right)
=\delta^\mathrm{gauge}_\alpha S\ .\label{Gausslaw}
 \end{align}
Here,
\begin{align}
\delta^{\rm gauge}_\alpha A^a_\mu
:=\nabb_\mu\alpha^a\equiv\nabla_\mu\alpha^a+{f^a}_{bc}
A^b_\mu\alpha^c \ ,~~
\delta^{\rm gauge}_\alpha \phi^I:=
i\alpha^a(T^a\phi)^I \ ,
\end{align}
denote an infinitesimal gauge transformation. 
{}Further, the momentum constraint (\ref{Momcon}) and
the Hamilton-Jacobi equations lead to
\begin{align}
0=&\int d^dx\sqrt{h}\,\epsilon^\mu\left(2\nabla^{\nu}\pi_{\mu\nu}
-\pi_I\nabb_\mu\,\phi^I-F^a_{\mu\nu}\pi^{a\nu}\right)
\nn\\
=&\int d^dx\left\{(\nabla_\mu\epsilon_\nu+\nabla_\nu\epsilon_\mu)
\ddif S{h_{\mu\nu}}+\epsilon^\mu \nabb_\mu\,\phi^I\ddif S{\phi^I}
+\epsilon^\mu F^a_{\mu\nu}\ddif S{A^a_\nu}\right\}
\nn\\
%
=&\delta_\epsilon S
%
-\int d^dx\sqrt{h}\,\epsilon^\mu A^a_\mu\left\{{\nabb^a}_{b\nu}\pi^{b\nu}
-i(T^a\phi)^I\pi_I\right\} \ .
\label{eAgauss}
\end{align}
Here,
\begin{align}
\delta_\epsilon\phi^I:=\cL_\epsilon\phi^I
\equiv\epsilon^\mu\partial_\mu\phi^I\ ,
\quad
\delta_\epsilon A^a_\mu:=\cL_\epsilon A^a_\mu\equiv\epsilon^\nu\partial_\nu A^a_\mu
+\partial_\mu\epsilon^\nu A^a_\nu \ ,
\quad
\delta_\epsilon h_{\mu\nu}:=\cL_\epsilon h_{\mu\nu}
\equiv\nabla_\mu\epsilon_\nu+\nabla_\nu\epsilon_\mu \ ,
\end{align}
are Lie derivatives with respect to $d$-dimensional diffeomorphism.
Noting that the second term in (\ref{eAgauss}) vanishes
because of (\ref{Gcon}) implies invariance of the on-shell action
under $d$-dimensional diffeomorphism.

{}For the purpose of solving the flow equation (\ref{SS}) using
systematic derivative expansions,
we divide the on-shell action into local and non-local parts:
\begin{align}
\frac1{2\kappa_{d+1}^2}S[h(x),\phi(x),A(x)]
\equiv\frac1{2\kappa_{d+1}^2}S_\mathrm{loc}
[h(x),\phi(x),A(x)]-\Gamma[h(x),\phi(x),A(x)]\ . 
\label{S:divided}
\end{align}
We next assign an additive
number called weight to each ingredient of the action as in a table below.
\begin{table}[h]
\begin{center}
\begin{tabular}{c|c}
elements&weight\\\hline
$h_{\mu\nu}(x),\phi^I(x),\Gamma[h,\phi,A]$&0\\\hline
$\partial_\mu,A^a_\mu(x)$&1\\\hline
$R,R_{\mu\nu},R_{\mu\nu\rho\sigma},\partial^2,\dots$&2\\\hline
$\ddif{}{A^a_\mu(x)}$&$d-1$\\\hline
$\ddif{}{h_{\mu\nu}(x)},\ddif{}{\phi^I(x)}$&$d$\\
\end{tabular}
\end{center}
\caption{assignment of weights}
\end{table}
The weight of the gauge field is assigned to be $w=1$ 
because of gauge invariance.
We parametrize the local Lagrangian as below
\begin{align}
 \cL_{\rm loc}=\sum_{w=0,2,4,\cdots}[\cL_{\rm loc}]_w \ ,
\end{align}
where 
\begin{align}
[\cL_{\rm loc}]_0=&W(\phi) \ ,
\\
[\cL_{\rm loc}]_2=&-\Phi(\phi)R_{(d)}+\frac12M_{IJ}(\phi)
\nabb^\mu\,\phi^I\nabb_\mu\,\phi^J\ .
\end{align}
It is important here that all the local terms are taken to
be gauge invariant.
We also define
\begin{align}
  S_{{\rm loc};w-d}:=\int d^dx\sqrt{h}\,[\cL_{\rm loc}]_w \ .
\end{align}
Note that $d^dx$ has a weight $-d$.

Inserting (\ref{S:divided}) into the flow equation and then 
decomposing it depending on weights, we find
for $w=0$
\begin{equation}
V(\phi)=-\frac d{4(d-1)}W^2(\phi)+\frac12L^{IJ}(\phi)\partial_IW(\phi)
		\partial_JW(\phi) \ .
\label{feq0}
\end{equation}
{}For $w=2$,
\begin{align}
-1=&\frac{d-2}{2(d-1)}W(\phi)\Phi(\phi)-L^{IJ}(\phi)\partial_IW(\phi)
		\partial_J\Phi(\phi) \ ,
\label{feq2-1}\\
\frac12L_{IJ}(\phi)=&-\frac{d-2}{4(d-1)}W(\phi)M_{IJ}(\phi)-L^{KL}(\phi)
\partial_KW(\phi)\Gamma_{L;IJ}(\phi)
\nonumber\\
&-W(\phi)\partial_I\partial_J\Phi(\phi)
-\frac1{2J(\phi)}M_{IK}(\phi)
M_{JL}(\phi)(T^a\phi)^K(T^a\phi)^L\ ,
\label{feq2-2}\\
0=&W(\phi)\partial_K\Phi(\phi)+L^{IJ}(\phi)\partial_IW(\phi)M_{JK}(\phi)\ .	\label{feq2-3}
\end{align}
{}For $w=4$, 
\begin{equation}
\frac14J(\phi)F^a_{\mu\nu}F^{a\mu\nu}=[\{S,S\}]_4\ .
\label{feq4}
\end{equation}
{}For $w=d$ with $d\neq 4$, 
\begin{align}
[\mathcal L_d]_d=&\frac{2\kappa_{d+1}^2W(\phi)}{2(d-1)}
\frac2{\sqrt{h}}h_{\mu\nu}\ddif\Gamma{h_{\mu\nu}}-\frac{2\kappa_{d+1}^2}
{\sqrt{h}}L^{IJ}(\phi)\partial_IW(\phi)\ddif\Gamma{\phi^J}\nonumber\\
&-\frac{2\kappa_{d+1}^2}{hJ(\phi)}h_{\mu\nu}
\ddif{S_{\mathrm{loc};2-d}}{A^a_\mu}\ddif\Gamma{A^a_\nu}+[\{S_\mathrm{loc},
S_\mathrm{loc}\}]_d\ .
\label{feqd}
\end{align}

In the AdS/CFT correspondence,
$h_{\mu\nu}(x), \phi^I(x)$ and $A_\mu^a(x)$ are identified with
a background metric, a coupling function associated with an
gauge invariant operator $O_I$, and a background gauge potential
of $G$, respectively, in the boundary QFT.
Then, we see that
(\ref{feqd}) is equivalent to
the local RG equation,
which
specifies how the coupling functions $\phi^I(x)$ and $A_\mu^a(x)$ flow under
local Weyl transformation \cite{O1991}.
In particular, by rewriting (\ref{feqd}) as
\begin{align}
2h_{\mu\nu}\frac{\delta\Gamma}{\,\,\delta h_{\mu\nu}}
&-\frac{2(d-1)}{W}L^{IJ}\partial_IW
\frac{\delta\Gamma}{\delta\phi^J}
-\frac{2(d-1)}{W}\frac{1}{J}h_{\mu\nu}
\frac{1}{\sqrt{h}}\,\frac{\delta S_{{\rm loc};2-d}}{\delta A_\mu^a}
\frac{\delta \Gamma}{\delta A_\nu^a}
\nn\\
&=\frac{1}{2\kappa_{d+1}^2}\frac{2(d-1)}{W}
\sqrt{h}\,\bigg(
[\cL_d]_d-\left[\{S_{\rm loc},S_{\rm loc}\}\right]_d
\bigg)
\ ,
\label{localRG}
\end{align}
the scalar $\beta$-function
associated with a coupling function $\phi^I$ 
reads
\begin{equation}
\beta^I(\phi):=-\frac{2(d-1)}{W(\phi)}L^{IJ}(\phi)\partial_JW(\phi) \ ,
\label{beta:grad}
\end{equation}
where $\partial_I=\partial/\partial\phi^I$.
Using (\ref{feq2-3}), this can be recast as
\begin{equation}
\beta^I(\phi)=+2(d-1)M^{IJ}(\phi)\partial_J\Phi(\phi) \ ,
\end{equation}
with $M^{IJ}=M_{IJ}^{-1}$.

The coefficient of
$\delta\Gamma/\delta A^a_\mu$ defines a vector $\beta$-function as
\begin{equation}
\beta^a_\mu(\phi,A):=
-\frac1{\sqrt h}\frac{2(d-1)}{W(\phi)J(\phi)}h_{\mu\nu}
	\ddif{S_{\text{loc;}2-d}}{A^a_\nu}
=
+i\frac{2(d-1)}{W(\phi)J(\phi)}
M_{IJ}(\phi)(T^a\phi)^I\nabb_\mu\phi^J\ .
\label{vecbeta}
\end{equation}
{}Following \cite{JO}, we define 
$\beta_\mu^a\equiv\rho_I^a\,\nabb_\mu\phi^I$ so that
\begin{align}
  \rho_I^a=+i\frac{2(d-1)}{W(\phi)J(\phi)}
M_{IJ}(\phi)(T^a\phi)^J \ .
\end{align}

We now show that $W, L_{IJ}$ and $V$ can be expressed in
terms of $\Phi$ and $M_{IJ}$.
{}From (\ref{feq2-1}) and (\ref{feq2-3}), we obtain
\begin{align}
  -W^{-1}=M^{IJ}\partial_I\Phi\,\partial_J\Phi
+\frac{d-2}{2(d-1)}\,\Phi \ .
\label{solW}
\end{align}
Next, using (\ref{feq2-3}), (\ref{feq2-2}) becomes
\begin{align}
  L_{IJ}
=2W
\left[-\frac{d-2}{4(d-1)}\,M_{IJ}
-D_ID_J\Phi
-\frac1{2JW}M_{IK}
M_{JL}(T^a\phi)^K(T^a\phi)^L
\right] \ .
\label{solL}
\end{align}
Here, $\Gamma^I_{JK}$ is the Levi-Civita connection
with respect to $M_{IJ}$, and $D_I$ is a covariant derivative defined with the
connection. {}For a consistency check of (\ref{solL}),
we rewrite (\ref{feq2-3}) as
\begin{align}
 -W^{-2}\,\partial_IW=
W^{-1}\,L_{IJ}\,M^{JK}\,\partial_K\Phi \ .
\end{align}
Using (\ref{solW}) and (\ref{solL}), the RHS takes the form
\begin{align}
-&\frac{d-2}{2(d-1)}\,\partial_I\Phi
-2D_ID_J\Phi\,\partial^J\Phi
-\frac{1}{JW}M_{IK}(T^a\phi)^K\,(T^a\phi)^L\,\partial_L\Phi
\nn\\
&=
\partial_I\left(
-\frac{d-2}{2(d-1)}\,\Phi
-M^{JK}\,\partial_J\Phi\,\partial_K\Phi
\right)
-\frac{1}{JW}M_{IK}(T^a\phi)^K\,(T^a\phi)^L\,\partial_L\Phi
\ .
\end{align}
{}Because $\Phi(\phi)$ is gauge invariant by definition, we have
\begin{align}
  (T^a\phi)^I\,\partial_I\Phi=0 \ .
\label{delPhi=0}
\end{align}
This ensures 
that (\ref{solL}) is consistent indeed.
{}Finally, (\ref{feq0}) and (\ref{feq2-3}) 
gives
\begin{align}
  V
&=-W^3\left[
\partial^I\Phi\,\partial^J\Phi\left(
D_ID_J\Phi-\frac{1}{2(d-1)}M_{IJ}
\right)
-\frac{d(d-2)}{8(d-1)^2}\,\Phi
\right] \ .
\end{align}

Vacuum expectation value of the stress tensor 
in the
presence of the background fields $h,\phi$ and $A$ is defined by
\begin{align}
\langle T^{\mu\nu}(x)\rangle:=
\frac2{\sqrt{h}}\ddif{\Gamma
	[h,\phi,A]}{h_{\mu\nu}(x)} \ . 
\end{align}
It follows from (\ref{localRG}) that the trace of the 
stress tensor becomes
\begin{align}
\left\langle{T^\mu}_\mu\right\rangle
=&\frac{2(d-1)}{2\kappa_{d+1}^2W(\phi)}
\bigg([\mathcal L_d]_d
-[\{S_\mathrm{loc},S_\mathrm{loc}\}]_d
\bigg)
-\beta^I(\phi)\frac1{\sqrt{h}}\ddif\Gamma{\phi^I}
-\beta_\mu^a\frac{1}{\sqrt{h}}
\ddif\Gamma{A^a_\mu} \ .
\label{traceT}
\end{align}
As explained in \cite{FMS1, FS}, the flow equation cannot 
determine $S_{{\rm loc};0}$ uniquely, reflecting
an ambiguity of adding local counterterms to $\Gamma$.

In a computation of the Weyl anomaly, $S_{{\rm loc};0}$ 
is manifested as
a degree of freedom of adding a total derivative \cite{FS}.
To see this, we note that under an infinitesimal Weyl transformation,
$S_{{\rm loc};0}$ transforms as
\begin{align}
\int d^dx\, 2\,\sigma(x)\,h_{\mu\nu}(x)
\frac{\delta}{\delta h_{\mu\nu}(x)}\,S_{{\rm loc};0}
=\int d^dx\sqrt{h}\,
\partial_\mu\sigma\cJ_d^\mu \ ,
\end{align}
because $S_{{\rm loc};0}$ is invariant under global
scale transformations. Thus,
\begin{align}
  2\,h_{\mu\nu}(x)
\frac{\delta}{\delta h_{\mu\nu}(x)}\,S_{{\rm loc};0}
=-\sqrt{h}\,\nabla_\mu\cJ_d^\mu \ .
\end{align}
{}From this relation, we obtain
\begin{align}
\left\langle{T^\mu}_\mu\right\rangle
=&\frac{2(d-1)}{2\kappa_{d+1}^2W(\phi)}
\bigg([\mathcal L_d]_d
-[\{S_\mathrm{loc},S_\mathrm{loc}\}']_d
\bigg)
-\frac{1}{2\kappa_{d+1}^2}\nabla_\mu\cJ^\mu_{d}
\nn\\
&-\beta^I(\phi)\frac1{\sqrt{h}}
\frac{\delta}{\delta\phi^I}
\left(
\Gamma-\frac{1}{2\kappa_{d+1}^2}S_{{\rm loc};0}
\right)
-\beta_\mu^a(\phi)\frac1{\sqrt{h}}
\frac{\delta}{\delta A_\mu^a}
\left(
\Gamma-\frac{1}{2\kappa_{d+1}^2}S_{{\rm loc};0}
\right) \ .
\end{align}
Here, $\{S_\mathrm{loc},S_\mathrm{loc}\}'$
denotes the bracket $\{S_\mathrm{loc},S_\mathrm{loc}\}$ with 
$[\mathcal L_\mathrm{loc}]_d$ removed from $S_{\rm loc}$.
Hence, 
the Weyl anomaly, which is defined as
\begin{align}
\mathcal W_d(x):=\left\langle{T^\mu}_\mu(x)\right\rangle\Big|_{\beta=0}
\ ,
\end{align}
contains a total derivative that comes from
$S_{{\rm loc};0}$.
It is important to note that in the AdS/CFT 
correspondence, $\cJ_d^\mu$ is the only origin of the Virial
current, which might spoil conformal symmetry of scale invariant
field theories. For an excellent review on relations
between scale and conformal symmetry, see \cite{Nak;review}.
As an operator, $\cJ_d^\mu$ is proportional to an identity
operator, and therefore gives no obstacle to having CFTs.
This is natural because we are working on QFTs with gravity duals.
Effects of the Virial current here are only manifested as
ambiguities of local counterterms addded to $\Gamma$.
{}For the moment, we work in the scheme 
$S_{\rm{loc};0}=0$ because this is simple and natural 
in the flow equation.
{}For a discussion on how $S_{{\rm loc};0}$ affects
the coefficients in $\langle T^\mu_\mu\rangle$, see \cite{Kal}.
Another choice of the scheme
will be discussed below 
for the purpose of studying a holographic $c$-theorem.

An analysis of the Gauss's law constraint is straightforward.
With the HJ equations, (\ref{Gcon}) can be regarded as a constraint
on the on-shell action:
\begin{align}
  {\nabb^a}_{b\mu}\frac{\delta S}{\delta A^b_\mu}
-i(T^a\phi)^I\,\frac{\delta S}{\delta \phi^I}=0 \ .
\label{gauss;S}
\end{align}
Inserting (\ref{S:divided}) into this, 
we see that the local terms give no contribution
because they are gauge invariant by definition.
Then, (\ref{gauss;S}) reduces to 
\begin{align}
  {\nabb^a}_{b\mu}\frac{\delta \Gamma}{\delta A^b_\mu}
-i(T^a\phi)^I\,\frac{\delta \Gamma}{\delta \phi^I}=0 \ .
\end{align}
Because the vev's of the gauge invariant
operators and currents in the presence of the background
fields are given by
\begin{equation}
\langle O_I(x)\rangle:=\frac1{\sqrt h}\ddif\Gamma{\phi^I(x)}\ ,
~~
\langle J^{a\mu}(x)\rangle:=\frac1{\sqrt h}\ddif\Gamma{A^a_\mu(x)}
\ ,
\end{equation}
we obtain an
operator identity 
\begin{equation}
\nabb_\mu J^{a\mu}=i(T^a\phi)^IO_I\ .\label{opid}
\end{equation}

We now give some comments on properties of the vector 
$\beta$-function that hold for $d$-dimensional QFTs with gravity duals. 
These were first obtained in \cite{N13}, and
the rest of this section may be regarded as a review
of part of that paper.
{}First, we have already observed that (\ref{vecbeta})
exhibits the gradient property. 
Second, an orthogonal relation between scalar and
vector $\beta$-functions is easy to verify in the AdS/CFT correspondence
thanks to the gauge invariance (\ref{delPhi=0}):
\begin{align}
\rho^a_I\beta^I=
i\frac{4(d-1)^2}{JW}\,(T^a\phi)^I\partial_I\Phi=0\ .\label{ortho}
\end{align}
In addition, anomalous dimensions receive non-trivial contributions 
from operator mixing: differentiating the local RG equation 
\begin{align}
(\text{local terms})
=\int d^dx\Bigg\{2h_{\mu\nu}(x)\ddif{}{h_{\mu\nu}(x)}+\beta^I
[\phi(x)]\ddif{}{\phi^I(x)}
+\rho^a_I[\phi(x)]\nabb_\mu\phi^I(x)\ddif{}{A^a_\mu(x)}
\Bigg\}\Gamma[\phi,h,A] \ ,
\end{align}
with $\phi^I$ and $A_\mu^a$,
we obtain the RG equations of correlation functions of
$O_I$ and $J^{a\mu}$, of which
the anomalous dimensions of $O_I$ and $J^{a\mu}$ are
read off as
\begin{align}
{\gamma^I}_J=-\partial_J\beta^I+i\rho^a_J(T^a\phi)^I
\ ,~~
%
{\gamma^a}_b=i\rho^c_I\delta_{bc}(T^a\phi)^I \ ,
\end{align}
respectively. Here we employ the operator identity (\ref{opid}).
It then follows
\begin{align}
{\gamma^I}_J
&=-2(d-1)\partial_J\partial^I\Phi(\phi)-\frac{2(d-1)}{W(\phi)J(\phi)}
M_{JK}(\phi)(T^a\phi)^K(T^a\phi)^I\ ,
\\
{\gamma^a}_b
&=-\frac{2(d-1)}{W(\phi)J(\phi)}M_{IJ}(\phi)(T^a\phi)^I\delta_{bc}
(T^c\phi)^J\ .
\end{align}
These expressions exhibit the suggested Higgs-like relation 
manifestly. 
{}Finally, the equivalence
\begin{equation}
	\beta^a_\mu=0\iff\nabb_\mu J^{a\mu}=0\label{nonreno}
\end{equation}
can also be shown. 
Recalling (\ref{opid}), $\Leftarrow$ is obviously true because the
conservation of the current implies $(T^a\phi)^I=0$. On the other hand, if
$\beta^a_\mu=0$, we have two possibilities: (i) $(T^a\phi)^I=0$ and (ii)
$\nabb_\mu\phi^I=0$ because we are assuming $M_{IJ}$ to be invertible (and
$d\neq1$). In case of (i), we have the current conservation via the 
operator identity (\ref{opid}).
In case of (ii), since $\phi^I$ does not belong to a singlet, we must have
$\phi^I=\text{const.}=0$, and this again results in the current conservation.

\section{Explicit calculations in four dimensions}

{}For $d=4$, the equation (\ref{feq4}) should not be imposed, and
for $w=4$ the flow equation (\ref{floweq}) yields
the local RG equation (\ref{localRG}) instead.
Using the formulae given in appendix \ref{formulae}
together with (\ref{traceT}),
we arrive at
the explicit expression of the trace of the stress tensor:
\begin{align}
\langle{T^\mu}_\mu\rangle=&
\frac6{2\kappa_5^2W(\phi)}\frac14J(\phi)F^a_{\mu\nu}F^{a\mu\nu}
-\beta^I(\phi)\frac1{\sqrt{h}}\ddif\Gamma{\phi^I}
-\beta^a_\mu(\phi,A)\frac1{\sqrt{h}}\ddif\Gamma{A^a_\mu}
\nn\\
&-\frac{6\Phi^2}{2\kappa_5^2W}R^{\mu\nu}R_{\mu\nu}
+\left(\frac{2\Phi^2}{2\kappa_5^2W}
-\frac3{2\kappa_5^2W}L^{IJ}\partial_I\Phi\partial_J\Phi\right)
R_{(4)}^2
\nn\\
&+\frac{12\Phi}{2\kappa_5^2W}\partial_I\Phi\cdot E^{\mu\nu}
\nabb_\mu \nabb_\nu\,\phi^I
-\frac6{2\kappa_5^2W}L^{JK}\partial_{(J}\Phi M_{K)I}\cdot
R_{(4)}\nabb^2\phi^I
\nn\\
&+\frac{6\Phi}{2\kappa_5^2W}(2\partial_I\partial_J\Phi
+M_{IJ})E^{\mu\nu}\nabb_\mu\,\phi^I \nabb_\nu\,\phi^J
\nn\\
&+\frac1{2\kappa_5^2W}\left(\Phi M_{IJ}
-6L^{KL}\partial_{(K}\Phi\Gamma_{L);IJ}\right)\cdot 
R_{(4)}\nabb^\mu\,\phi^I\nabb_\mu\,\phi^J
\nn\\
&+\left(\frac6{2\kappa_5^2W}\partial_I\Phi\partial_J\Phi
-\frac3{2\kappa_5^2W}L^{KL}M_{IK}M_{JL}\right)\nabb^2\phi^I\nabb^2\phi^J
\nn\\
&+\left(\frac{12}{2\kappa_5^2W}\partial_I\Phi\partial_J\partial_K\Phi
+\frac3{2\kappa_5^2W}\partial_I\Phi M_{JK}
-\frac6{2\kappa_5^2W}L^{LM}M_{I(L}\Gamma_{M);JK}\right)
\nabb^2\phi^I\nabb^\mu\,\phi^J\nabb_\mu\,\phi^K
\nn\\
&-\frac6{2\kappa_5^2W}\partial_I\Phi\partial_J\Phi\cdot 
\nabb^\mu \nabb^\nu\,\phi^I\nabb_\mu \nabb_\nu\,\phi^J
\nn\\
&-\frac6{2\kappa_5^2W}\partial_I\Phi\left(2\partial_J\partial_K\Phi
+M_{JK}\right)\cdot \nabb^\mu \nabb^\nu\,\phi^I
\nabb_\mu\,\phi^J\nabb_\nu\,\phi^K
\nn\\
&+\Bigg(\frac1{4\kappa_5^2W}M_{IJ}M_{KL}-\frac3{4\kappa_5^2W}M_{IK}M_{JL}
		+\frac3{4\kappa_5^2W}(M_{IJ}\partial_K\partial_L\Phi+M_{KL}
			\partial_I\partial_J\Phi)
\nn\\
&\hspace{50pt}-\frac3{2\kappa_5^2W}(M_{IK}\partial_J\partial_L\Phi+M_{JL}\partial_I
		\partial_K\Phi)+\frac6{2\kappa_5^2W}(\partial_I\partial_J\Phi\partial_K\partial_L\Phi-\partial_I\partial_K\Phi\partial_J\partial_L\Phi)
\nn\\
&\hspace{50pt}-\frac3{2\kappa_5^2W}L^{MN}\Gamma_{M;IJ}\Gamma_{N;KL}\Bigg)\nabb^\mu\,\phi^I\nabb_\mu\,\phi^J
\nabb^\nu\,\phi^K\nabb_\nu\,\phi^L
\ .
\label{explicitT}
\end{align}

Now we compare these results with those
in \cite{JO}, where a generic form of 
$\langle T_\mu^\mu\rangle$ is given in accord with
symmetry constraints.
{}For details, see Appendix \ref{JO:trace}.
(\ref{1st}), (\ref{2nd}), (\ref{3rd}) and (\ref{4th})
are easily solved as
\begin{equation}
A=-\frac{12\Phi^2}{2\kappa_5^2W}\ ,
\quad 
C=-\frac{3\Phi^2}{2\kappa_5^2W}\ ,
\quad
B=\frac{216}{2\kappa_5^2W}L^{IJ}\partial_I\Phi\partial_J\Phi\ 
,\label{ACB}
\end{equation}
\begin{equation}
W_I=\frac6{2\kappa_5^2W}\Phi\partial_I\Phi=\frac3{2\kappa_5^2W}\partial_I\Phi^2.
\label{WI}
\end{equation}
Using (\ref{WI}), (\ref{5th}) yields
\begin{equation}
	G_{IJ}=-\frac{12}{2\kappa_5^2W^2}\Phi\partial_{(I}W\partial_{J)}\Phi+\frac6
		{2\kappa_5^2W}\partial_I\Phi\partial_J\Phi-\frac6{2\kappa_5^2W}\Phi M_{IJ}.
\label{GIJ}
\end{equation}
{}From (\ref{6th}), (\ref{7th}) and (\ref{8th}), we find
\begin{align}
H_I&=0\ ,
\\
E_I&=\frac{36}{2\kappa_5^2W}L^{JK}\partial_{(J}\Phi M_{K)I}\ ,
\\
F_{IJ}&=-\frac{18}{2\kappa_5^2W}\partial_I\Phi\partial_J\Phi
-\frac6{2\kappa_5^2W}\Phi M_{IJ}
+\frac{36}{2\kappa_5^2W}L^{KL}\partial_{(K}\Phi\Gamma_{L);IJ}\ .
\end{align}
(\ref{10th}) and (\ref{11th}) give
\begin{equation}
S_{IJ}=S_{(IJ)}=V_{IJ}=\frac3{2\kappa_5^2W}\partial_I\Phi\partial_J\Phi
\ ,
\end{equation}
from which (\ref{9th}) leads to
\begin{equation}
A_{IJ}=-\frac6{2\kappa_5^2W}L^{KL}M_{IK}M_{JL}\ .
\label{AIJ}
\end{equation}
The rest of the equations given in Appendix \ref{JO:trace}
requires hard work to solve. 
However, since there are six
equations left and six coefficient functions to be determined, viz.
$B_{IJK},T_{IJK}$, $C_{IJKL},\beta_f,Q_I$ and $P_{IJ}$, thus it is expected that there is a
solution.

It is argued in \cite{JO} that the quantities appearing in 
$\langle T^\mu_\mu\rangle$ must satisfy integrability conditions,
that is, Wess-Zumino consistency conditions associated
with local RG transformations:
\begin{align}
  \left[ \Delta_\sigma,\Delta_{\sigma'}\right]\Gamma=0 \ .
\end{align}
Here
\begin{align}
  \Delta_\sigma:=\int d^dx\,\sigma(x)\left(
2h_{\mu\nu}\frac{\delta}{\,\,\delta h_{\mu\nu}}
+\beta^I
\frac{\delta}{\delta\phi^J}
+\beta_\mu^a\,
  \frac{\delta}{\delta A_\mu^a}
\right) \ .
\end{align}
As discussed before, the AdS/CFT correspondence always 
gives us trivial Virial currents.
Then, no nontrivial modification of the scalar
and vector $\beta$-functions arises that is necessary 
for gauge invariance of the $\beta$ functions.
Therefore, we can show that a number of integrability conditions derived in
\cite{JO} still hold on their own. We also obtain additional integrability
conditions concerned with the external gauge field. 
To list some of the integrability conditions that
play an important role in this paper, we have
\begin{align}
  \partial_IA&=G_{IJ}\beta^J-\cL_\beta W_I \ ,
\label{WZ;delA}
\\
  \rho^a_I\beta^I&=0 \ .\label{WZ;ortho}
\end{align}
Here $\cL_\beta$ denotes a Lie derivative associated with
the vector $\beta^I$, which acts on $W_I$ as
\begin{align}
  \cL_\beta W_I\equiv\beta^K\partial_K W_I+\partial_I\beta^KW_K \ .
\end{align}
The coefficients given in 
(\ref{ACB})-(\ref{AIJ}) should satisfy all the integrability
conditions, because 
the flow equation is formulated on the basis that
the effective action $\Gamma$ does exist once a bulk
gravity model is given.
In fact, a straightforward computation shows that
(\ref{WZ;delA}) holds indeed.
{}Furthermore, 
(\ref{WZ;ortho}) is nothing but the orthogonality 
condition (\ref{ortho}), 
which is already verified from the gauge invariance
of $\Phi$.

We end this paper by making some comments on
the $c$-theorem of RG flows in higher dimensions and a holographic
$c$-function.
As shown by Jack and Osborn in \cite{JO}, 
(\ref{WZ;delA}) can be rewritten as
\begin{align}
\beta^I\partial_I\tilde{A}=G_{IJ}\beta^I\beta^J \ ,
\label{delAtilde}
\end{align}
with 
\begin{align}
 \tilde{A}:=A+W_I\beta^I \ .
\end{align}
Proving positive-definiteness of $G_{IJ}$, if possible, implies
that $\tilde{A}$ decreases monotonically under
RG flows. 
$A$ and $G_{IJ}$ in (\ref{ACB}) and (\ref{GIJ}) 
are obtained in the scheme $S_{\rm loc;0}=0$.
As evident, this $G_{IJ}$ is not positive definite even if
$L_{IJ}$ is assumed to be so.
However, $G_{IJ}$ takes a different
expression by working in 
a different scheme with $S_{\rm loc;0}\ne 0$.
In fact, it is argued in \cite{JO} that adding local counterterms 
to $\Gamma$ gives rise to a shift
\begin{align}
\tilde A\to \tilde A':=\tilde A+g_{IJ}\beta^I\beta^J \ ,~~
G_{IJ}\to G'_{IJ}:=G_{IJ}+\cL_\beta g_{IJ} \ ,
\label{shift}
\end{align}
which leaves (\ref{delAtilde}) unchanged.
Here, we show that an appropriate choice of $g_{IJ}$ 
maps $\tilde A$ to a holographic $c$-function.
Relations between a holographic $c$-function
and $\tilde A$ were first studied in \cite{Erd01}.
Our aim in this paper is to make a full identification
of schemes
where a holographic $c$-function is related directly
with an trace anomaly coefficient.
As discussed in \cite{Gir98,Free99}, 
a holographic $c$-function for $d=4$ is defined as
\begin{align}
  c_h:=-\frac{27}{2\kappa_5^2}\,\frac{1}{W^3} \ .
\label{ch}
\end{align}
Here the overall factor is chosen so that the value of
$c_h$ at a fixed point equals that of $C$ given
in (\ref{ACB}). 
It follows from (\ref{ch}), (\ref{solW}) and (\ref{solL})
that
\begin{align}
  \beta^I\partial_Ic_h=\frac{1}{2}\,c_h\,L_{IJ}\,\beta^I
\,\beta^J\ .
\label{del;ch}
\end{align}
This relation was first derived in \cite{Ans00}, although
we prove it when $\phi^I$ is promoted
to space-time dependent couplings. 
The gradient flow nature becomes more manifest by rewriting
the scalar $\beta$-function (\ref{beta:grad}) as
\begin{align}
 \beta^I=\frac{2}{c_h}L^{IJ}\partial_J c_h \ .
\end{align}
The positivity of
$c_h$ together with positive definiteness of $L_{IJ}$ 
guarantees that $c_h$ is indeed a monotonically decreasing
function. $c_h L_{IJ}$ is to be identified with a Zamolodchikov
metric.

{}For the purpose of relating $\tilde A'$ to $c_h$,
we take $g_{IJ}$ to the most general form:
\begin{align}
  2\kappa_5^2\,g_{IJ}:=X(\phi)\,\partial_I\Phi\,\partial_J\Phi
+Y(\phi)\,M_{IJ}
\ ,
\label{gIJ}
\end{align}
where
\begin{align}
  X(\phi)=&x_1(\Phi)\left(\partial\Phi\cdot\partial\Phi\right)
+x_2(\Phi) \ ,
\nn\\
Y(\phi)=&(3-x_1(\Phi))\left(\partial\Phi\cdot\partial\Phi\right)^2
+(4\Phi-x_2(\Phi))\left(\partial\Phi\cdot\partial\Phi\right)
+\Phi^2 \ ,
\end{align}
with $x_1,x_2$ being arbitrary functions of $\Phi=\Phi(\phi^I)$
and $(\partial\Phi\cdot\partial\Phi)=M^{IJ}\partial_I\Phi\partial_J\Phi$.
{}From this mapping, it turns out that $\tilde A'=4c_h$.
{}Furthermore, we can easily show that
\begin{align}
  G'_{IJ}\beta^I\beta^J
=2c_h\,L_{IJ}\beta^I\beta^J \ .
\end{align}
This implies that
(\ref{delAtilde}) after a shift
(\ref{shift}) with (\ref{gIJ})
is identical with (\ref{del;ch}).

\section*{Acknowledgments}
We would like to thank Yu Nakayama for many useful discussions.

\makeatletter
\renewcommand{\theequation}
{\Alph{section}.\arabic{equation}}
\@addtoreset{equation}{subsection}
\makeatother
\appendix

\section{Notations}
\label{notation}

Let $\phi^I$ be a charged scalar.
We divide the index $I$ into two parts: $I=(i,\alpha_i)$.
For each $i$, the charged field transforms 
as a representation $R_i$ under the 
gauge group $G$.
$\alpha_i=1,2,\cdots,{\rm dim}R_i$ is an index of $R_i$.
The generator of $G$, $(T^a)^I_{~J}$, defined in this paper
refers to
\begin{align}
  (T^a)^I_{~J}=\delta^i_j\,(t^a_{(i)})^{\alpha_i\beta_i} \ .
\end{align}
with $I=(i,\alpha_i),\,J=(j,\beta_j)$.
$t^a_{(i)}$ is the generators of $G$ that belong to
the representation $R_i$.
The covariant derivative $\nabb$ acts on $\phi^I$ as
\begin{align}
  \nabb_\mu\phi^I=
  \nabb_\mu\phi^{i,\alpha_i}
:=&\nabla_\mu\phi^{i,\alpha_i}
-iA_\mu^a\,(T^a)^I_{~J}\phi^J
\nn\\
=&\nabla_\mu\phi^{i,\alpha_i}
-iA_\mu^a\,\sum_{\beta_i}(t^a_{(i)})^{\alpha_i\beta_i}
\phi^{i,\beta_i} \ .
\end{align}

In addition, we define a symbol $(~)$ to denote symmetric parts of tensors:
\begin{equation}
	2X_{(IJ)}:=X_{IJ}+X_{JI},
\end{equation}
and denote the Levi-Civita connection in the theory space constructed from
$M_{IJ}$ $\Gamma_{I;JK}$, i.e.
\[ \Gamma_{I;JK}:=\frac12\left(\partial_JM_{IK}+\partial_KM_{IJ}-\partial_IM_{JK}\right).
	\]

\section{Some Useful Formulae}
\label{formulae}

The following formulae are useful and valid for any $d$:
\begin{align}
\ddif{}{h_{\mu\nu}}S_{\mathrm{loc};-d}&=
\frac12\sqrt{h}\,h^{\mu\nu}W(\phi) \ ,
\\
\ddif{}{\phi^I}S_{\mathrm{loc};-d}=&\sqrt{h}\,\partial_IW(\phi) \ ,
\end{align}

\begin{align}
	\ddif{}{h_{\mu\nu}}S_{\mathrm{loc};2-d}&=
\sqrt{h}\,\Bigg\{\Phi(\phi)(R^{\mu\nu}
-\frac12h^{\mu\nu}R_{(d)})
-\nabb^{\mu}\nabb^{\nu}\Phi(\phi)
+h^{\mu\nu}\nabb^{2}\Phi(\phi)
\nn\\
&~~~~~~~~~+\frac12M_{IJ}(\phi)\left[\frac12h^{\mu\nu}\nabb^\rho
\phi^I\nabb_\rho\phi^J
-\nabb^\mu\phi^I\nabb^\nu\phi^J\right]\Bigg\} \ ,\\
\ddif{}{\phi^I}S_{\mathrm{loc};2-d}=&
\sqrt{h}\,\Big\{-\partial_I\Phi(\phi)R_{(d)}
-\Gamma_{I;JK}(\phi)\nabb^\mu\phi^J\nabb_\mu\phi^K-M_{IJ}(\phi)
\nabb^{2}\phi^J\Big\} \ .
\end{align}

{}From these results, we find
\begin{align*}
[\{&S_{\mathrm{loc};2-d},S_{\mathrm{loc};2-d}\}]_4\\
=&R^{\mu\nu}R_{\mu\nu}\cdot\Phi^2+R_{(d)}^2\left(-\frac d{4(d-1)}\Phi^2
+\frac12L^{IJ}\partial_I\Phi\partial_J\Phi\right)\\
&+E^{\mu\nu}\nabb_\mu \nabb_\nu\phi^I(-2\Phi\partial_I\Phi)\\
&+E^{\mu\nu}\nabb_\mu\phi^I\nabb_\nu\phi^J[-\Phi(2\partial_I\partial_J\Phi+M_{IJ})]\\
&+R_{(d)}\nabb^2\phi^I\Big[L^{JK}\partial_{(J}\Phi M_{K)I}\Big]\\
&+R_{(d)}\nabb^\mu\phi^I\nabb_\mu\phi^J\left[-\frac{d-2}{4(d-1)}\Phi M_{IJ}+L^{KL}
		\partial_{(K}\Phi\Gamma_{L);IJ}\right]\\
&+\nabb^2\phi^I\nabb^2\phi^J\left[-\partial_I\Phi\partial_J\Phi+\frac12L^{KL}
		M_{IK}M_{JL}\right]\\
&+\nabb^2\phi^I\nabb^\mu\phi^J\nabb_\mu\phi^K\left[-2\partial_I\Phi\partial_J\partial_K\Phi
		-\frac12\partial_I\Phi M_{JK}+L^{LM}M_{I(L}\Gamma_{M);JK}\right]\\
&+\nabb^\mu \nabb^\nu\phi^I\nabb_\mu \nabb_\nu\phi^J(\partial_I\Phi\partial_J\Phi)\\
&+\nabb^\mu \nabb^\nu\phi^I\nabb_\mu \phi^J\nabb_\nu\phi^K\Big[\partial_I\Phi(2\partial_J
		\partial_K\Phi+M_{JK})\Big]\\
&+\nabb^\mu\phi^I\nabb_\mu\phi^J\nabb^\nu\phi^K\nabb_\nu\phi^L\Bigg[-\frac d{16(d-1)}
		M_{IJ}M_{KL}+\frac14M_{IK}M_{JL}-\frac14(M_{IJ}\partial_K\partial_L\Phi+M_{KL}
			\partial_I\partial_J\Phi)\\
	&\hspace{150pt}+\frac12(M_{IK}\partial_J\partial_L\Phi+M_{JL}\partial_I\partial_K
		\Phi)-(\partial_I\partial_J\Phi\partial_K\partial_L\Phi-\partial_I\partial_K\Phi
			\partial_J\partial_L\Phi)\\
	&\hspace{150pt}+\frac12L^{MN}\Gamma_{M;IJ}\Gamma_{N;KL}\Bigg]\ .
\end{align*}

\section{Trace of stress tensor defined in \cite{JO}
and its relation to that obtained from bulk gravity
}
\label{JO:trace}

In \cite{JO}, Jack and Osborn wrote down the explicit form
of the trace of the stress tensor as 
\begin{align}
\hspace{-20pt}\langle{T^\mu}_\mu\rangle
=&CW_{\mu\nu\rho\sigma}^2-\frac14AE_4-\frac1{72}BR_{(4)}^2-E^{\mu\nu}G_{IJ}
		\nabb_\mu\phi^I\nabb_\nu\phi^J
\nn\\
&-\frac16R_{(4)}\left(E_I\nabb^2\phi^I+F_{IJ}\nabb^\mu\phi^I\nabb_\mu\phi^J\right)
\nn\\
&+\frac12A_{IJ}\nabb^2\phi^I\nabb^2\phi^J+B_{IJK}\nabb^2\phi^I\nabb^\mu\phi^J\nabb_\mu\phi^K+\frac12C_{IJKL}\nabb^\mu\phi^I
\nabb_\mu\phi^J\nabb^\nu\phi^J\nabb_\nu\phi^K
\nn\\
&+\frac14(F^{\mu\nu}F_{\mu\nu})\beta_f+F^{\mu\nu}\cdot P_{IJ}\nabb_\mu\phi^I\nabb_\nu\phi^J
\nn\\
&+2\nabb_\mu\left(E^{\mu\nu}W_I\nabb_\nu\phi^I+\frac16R_{(4)}H_I
\nabb^\mu\phi^I+S_{IJ}\nabb^\mu\phi^I\nabb^2\phi^J+T_{IJK}
\nabb^\mu\phi^I\nabb^\nu\phi^J\nabb_\nu\phi^K
\right.
\nn\\
&\left.\qquad\qquad+F^{\mu\nu}\cdot Q_I\nabb_\nu\phi^I\right)
\nn\\
&-\nabb^2\left(\frac16R_{(4)}D+U_I\nabb^2\phi^I+V_{IJ}\nabb^\mu\phi^I
\nabb_\mu\phi^J\right)+(\text{terms proportional to $\beta$-functions}) \ .
\end{align}
Here the most general total derivative terms are added.
{}For the purpose of matching with those results computed 
from the bulk gravity, however,
it is sufficient to set $D=0=U_I$
because there is no term proportional to $\nabb^2R_{(4)}$ or 
$\nabb^4\phi^I$ there with $S_{{\rm loc};0}=0$.
It is then straightforward to verify

\begin{align}
\langle{T^\mu}_\mu\rangle
=&\left(C-\frac14A\right)R_{\mu\nu\rho\sigma}^2
		+(-2C+A)R_{\mu\nu}^2+\left(\frac C3-\frac A4-\frac B{72}\right)R_{(4)}^2
			\nn\\
	&+E^{\mu\nu}\nabb_\mu\nabb_\nu\phi^I\cdot2W_I+E^{\mu\nu}\nabb_\mu\phi^I\nabb_\nu\phi^J
		\left[-G_{IJ}+2\partial_{(I}W_{J)}-2S_{(IJ)}\right]\nn\\
	&+R_{(4)}\nabb^2\phi^I\left(-\frac16E_I+\frac13H_I\right)+R_{(4)}\nabb^\mu\phi^I\nabb_\mu
		\phi^J\left(-\frac16F_{IJ}+\frac13\partial_{(I}H_{J)}\right)\nn\\
	&+\nabb^\mu R_{(4)}\nabb_\mu\phi^I\cdot\frac13H_I+\nabb^2\phi^I\nabb^2\phi^J
		\left(\frac12A_{IJ}+2S_{(IJ)}\right)+\nabb^\mu\nabb^\nu\phi^I\nabb_\mu
			\nabb_\nu\phi^J(-2V_{IJ})
\nn\\
&+\nabb_\mu\phi^{i,\alpha_i}\nabb^2\nabb^\mu\phi^{j,\beta_j}\left[2(S_{ij})_{\alpha_i\beta_j}
			-2(V_{ij})_{\alpha_i\beta_j}\right]
\nn\\
	&+\nabb^2\phi^I\nabb^\mu\phi^J\nabb_\mu\phi^K[B_{IJK}+2\partial_{(J}S_{K)I}+2T_{IJK}-\partial_I
		V_{JK}]\nn\\
	&+\nabb^\mu\nabb^\nu\phi^I\nabb_\mu\phi^J\nabb_\nu\phi^K[4T_{JKI}-4\partial_JV_{KI}]
			\nn\\
	&+\nabb^\mu\phi^I\nabb_\mu\phi^J\nabb^\nu\phi^K\nabb_\nu\phi^L\left[\frac12C_{IJKL}
		+\partial_{(I}T_{J)KL}+\partial_{(K}T_{L)IJ}-\frac12\partial_I\partial_JV_{KL}
			-\frac12\partial_K\partial_LV_{IJ}\right]\nn\\
	&+{(F^{\mu\nu}F_{\mu\nu})^{\alpha_i}}_{\beta_i}\left[\frac14{(\beta_f)^{\beta_i}}_{\alpha_i}+\phi^{i,\beta_i}Q_{i,\alpha_i}\right]\nn\\
	&+\nabb_\mu\phi^{i,\alpha_i}\left[(P_{ij})_{\alpha_i\beta_j}+2(\partial_iQ_j)_{\alpha_i\beta_j}+2(S_{ij})_{\alpha_i\beta_j}\right]
			{(F^{\mu\nu})^{\beta_j}}_{\gamma_j}\nabb_\nu\phi^{j,\gamma_j}\nn\\
	&+\nabb_\mu\phi^{i,\alpha_i}{(\nabb_\nu F^{\nu\mu})^{\beta_j}}_{\gamma_j}\left[2\delta_i^j\delta^{\gamma_j}_{\alpha_i}Q_{j,\beta_j}-2(S_{ij})_{\alpha_i\beta_j}
		\phi^{j,\gamma_j}\right]
\nn\\
	&+(\text{terms proportional to $\beta$-functions}).
\label{traceT:JO}
\end{align}

Comparing the coefficients of operators appearing in (\ref{explicitT})
and those in (\ref{traceT:JO}) gives
\begin{align}
C-\frac14A&=0\ ,
\label{1st}\\
-2C+A&=-\frac{6\Phi^2}{2\kappa_5^2W}\ ,
\label{2nd}
\\
\frac C3-\frac A4-\frac B{72}&=\frac{2\Phi^2}{2\kappa_5^2W}
-\frac3{2\kappa_5^2W}L^{IJ}\partial_I\Phi\partial_J\Phi\ ,
\label{3rd}
\\
2W_I&=\frac{12}{2\kappa_5^2W}\Phi\partial_I\Phi\ ,
\label{4th}
\\
-G_{IJ}+2\partial_{(I}W_{J)}-2S_{(IJ)}&=\frac{12}{2\kappa_5^2W}\Phi\partial_I
\partial_J\Phi+\frac6{2\kappa_5^2W}\Phi M_{IJ},
\label{5th}\\
-\frac16E_I+\frac13H_I
&=-\frac6{2\kappa_5^2W}L^{JK}\partial_{(J}\Phi M_{K)I}\ ,
\label{6th}\\
-\frac16F_{IJ}+\frac13\partial_{(I}H_{J)}-S_{(IJ)}
&=\frac1{2\kappa_5^2W}\Phi M_{IJ}
-\frac6{2\kappa_5^2W}L^{KL}\partial_{(K}\Phi\Gamma_{L);IJ}\ ,
\label{7th}\\
H_I&=0\ ,
\label{8th}\\
\frac12A_{IJ}+2S_{(IJ)}
&=\frac6{2\kappa_5^2W}\partial_I\Phi\partial_J\Phi-\frac3
{2\kappa_5^2W}L^{KL}M_{IK}M_{JL}\ ,
\label{9th}\\
-2V_{IJ}&=-\frac6{2\kappa_5^2W}\partial_I\Phi\partial_J\Phi\ ,
\label{10th}\\
2(S_{ij})_{\alpha_i\beta_j}-2(V_{ij})_{\alpha_i\beta_j}&=0\ ,
\label{11th}\\
B_{IJK}+2\partial_{(J}S_{K)I}+2T_{IJK}-\partial_IV_{JK}&=\frac{12}{2\kappa_5^2W}
		\partial_I\Phi\partial_J\partial_K\Phi+\frac3{2\kappa_5^2W}\partial_I\Phi M_{JK}
			-\frac6{2\kappa_5^2W}L^{LM}M_{I(L}\Gamma_{M);JK},\\
	4T_{JKI}-4\partial_JV_{KI}&=-\frac{12}{2\kappa_5^2W}\partial_I\Phi
		\partial_J\partial_K\Phi-\frac6{2\kappa_5^2W}\partial_I\Phi M_{JK},\\
	&\hspace{-130pt}\frac12C_{IJKL}+\partial_{(I}T_{J)KL}+\partial_{(K}T_{L)IJ}
		-\frac12\partial_I\partial_JV_{KL}-\frac12\partial_K\partial_LV_{IJ}\nn\\
	&\hspace{-76pt}=\frac1{4\kappa_5^2W}M_{IJ}M_{KL}-\frac3{4\kappa_5^2W}M_{IK}
		M_{JL}+\frac3{4\kappa_5^2W}(M_{IJ}\partial_K\partial_L\Phi
			+M_{KL}\partial_I\partial_J\Phi)\nn\\
	&\hspace{-56pt}-\frac3{2\kappa_5^2W}(M_{IK}\partial_J\partial_L\Phi+M_{JL}
		\partial_I\partial_K\Phi)+\frac6{2\kappa_5^2W}(\partial_I\partial_J\Phi
			\partial_K\partial_L\Phi-\partial_I\partial_K\Phi\partial_J\partial_L\Phi)\nn\\
	&\hspace{-36pt}-\frac3{2\kappa_5^2W}L^{MN}\Gamma_{M;IJ}\Gamma_{N;KL},\\
	\frac14{(\beta_f)^{\beta_i}}_{\alpha_i}+\phi^{i,\beta_i}Q_{i,\alpha_i}&=\frac6{2\kappa_5^2W}\frac14B(\phi)
		{\delta^{\beta_i}}_{\alpha_i},\\
	(P_{ij})_{\alpha_i\beta_j}+2(\partial_iQ_j)_{\alpha_i\beta_j}+2(S_{ij})_{\alpha_i\beta_j}&=0,\\
	2\delta_i^j\delta^{\gamma_j}_{\alpha_i}Q_{j,\beta_j}-2(S_{ij})_{\alpha_i\beta_j}\phi^{j,\gamma_j}&=0.
\end{align}

\end{document}